\documentclass[preprint]{iucr}              % DO NOT DELETE THIS LINE
     \paperprodcode{a000000}      % Replace with production code if known
     \paperref{xx9999}            % Replace xx9999 with reference code if known
     \papertype{FA}               % Indicate type of article
     \paperlang{english}          % Can be english, french, german or russian
     \journalcode{J}              % Indicate the journal to which submitted

     \journalyr{2001}
     \journaliss{1}
     \journalvol{57}
     \journalfirstpage{000}
     \journallastpage{000}
     \journalreceived{0 XXXXXXX 0000}
     \journalaccepted{0 XXXXXXX 0000}
     \journalonline{0 XXXXXXX 0000}

\begin{document}                  % DO NOT DELETE THIS LINE

     %-------------------------------------------------------------------------
     % The introductory (header) part of the paper
     %-------------------------------------------------------------------------

     % The title of the paper. Use \shorttitle to indicate an abbreviated title
     % for use in running heads (you will need to uncomment it).

\title{Numerical computation of critical properties and atomic basins
  from 3D grid electron densities}
\shorttitle{Topological properties from 3D grids}

     % Authors' names and addresses. Use \cauthor for the main (contact) author.
     % Use \author for all other authors. Use \aff for authors' affiliations.
     % Use lower-case letters in square brackets to link authors to their
     % affiliations; if there is only one affiliation address, remove the [a].
\author[a]{Katan}{C.}
\author[a]{Rabiller}{P.}
\author[b]{Lecomte}{C.}
\author[a]{Guezo}{M.}
\author[a]{Oison}{V.}
\cauthor[b]{Souhassou}{M.}{souhas@lcm3b.uhp-nancy.fr}{}
% \author[b]{Forename}{Surname}

\aff[a]{GMCM-UMR CNRS 6626, Universit\'e Rennes 1, av. Gal. Leclerc,
  35042 Rennes, \country{France}}
\aff[b]{LCM3B-UMR CNRS 7036, Universit\'e Henri Poincar\'e Nancy 1,
  54506 Vand\oe{}uvre l\`es Nancy, \country{France} }

     % Use \shortauthor to indicate an abbreviated author list for use in
     % running heads (you will need to uncomment it).

\shortauthor{Katan, Rabiller, Souhassou}

\maketitle                        % DO NOT DELETE THIS LINE

\begin{synopsis}
InteGriTy, a software package to compute topological properties of electron
densities given on 3D grids.
\end{synopsis}

\begin{abstract}
InteGriTy is a software package that performs topological analysis
following AIM approach on electron densities given on 3D grids. Use of
tricubic interpolation is made to get the density, its gradient and
hessian matrix at any required position. Critical points and
integrated atomic properties have been derived from theoretical
densities calculated for the compounds NaCl and TTF-2,5Cl$_2$BQ, thus
covering the different kinds of chemical bonds: ionic, covalent,
hydrogen bonds and other intermolecular contacts. 
\end{abstract}

     %-------------------------------------------------------------------------
     % The main body of the paper
     %-------------------------------------------------------------------------
     % Now enter the text of the document in multiple \section's, \subsection's
     % and \subsubsection's as required.

\section{Introduction}

Nowadays, very accurate electron densities can be obtained both experimentally and 
theoretically. Experimental methods of recovering charge densities require high 
resolution X-ray diffraction measurements on single crystals which are analysed 
within the context of aspherical models. From the theoretical point of view, not 
only crystals but also molecules, clusters and surfaces can be tackled using either 
quantum chemistry techniques ranging from standard Hartree-Fock calculations up 
to extremely accurate configuration interaction methods or techniques based on the 
Density Functional Theory (DFT) which are increasingly used to perform ab-initio 
calculations.  Once the total electron density is known, it can be analyzed in 
details by means of its topological properties within the Quantum Theory of Atoms 
in Molecules (Bader, 1990; Bader, 1994). 
With such an analysis one can go beyond a purely 
qualitative description of the nature and strength of interatomic interactions. 
It can also be used to define interatomic surfaces inside which  
atomic charges and moments are integrated.

The topological features of the total electron density $\rho(\mathbf{r})$ can be 
characterized by analyzing its gradient vector field 
$\mathbf{\nabla}\rho(\mathbf{r})$. Critical points (CPs) are located at points 
$\mathbf{r}_{\mbox{\scriptsize CP}}$ where 
$\mathbf{\nabla}\rho(\mathbf{r}_{\mbox{\scriptsize CP}}) = \mathbf{0}$ 
and the nature of each CP is 
determined from the curvatures ($\lambda_1$, $\lambda_2$, $\lambda_3$) 
of the density at this point. The latter are obtained 
by diagonalyzing the Hessian matrix $H_{ij} = {\partial^2 \rho(\mathbf{r}) \over
\partial x_i \partial x_j} (i,j=1,2,3)$. The CPs are denoted by a pair of integers
$(\omega, \sigma)$ where $\omega$ is the number of non-zero
eigenvalues of the Hessian matrix $H(\mathbf{r})$ and $\sigma$ the sum
of the signs of the three eigenvalues. In a three dimensional stable structure,
four types of CPs can be found: $(3,-3)$ Peaks corresponding to local maxima
of $\rho(\mathbf{r})$ which occur at atomic nuclear positions and in rare cases
at so called non-nuclear attractors; $(3,-1)$ Passes, corresponding to saddle 
points where $\rho(\mathbf{r})$ is maximum in the plane defined by the axes
corresponding to the two negative curvatures and minimum in the third direction, 
such bond critical points are found between pairs of bonded atoms; $(3,+1)$ Pales
where $\rho(\mathbf{r})$ is minimum in the plane defined by the axes associated
with the two positive
curvatures and maximum in the third direction, such ring critical points are
found within rings of bonded atoms; $(3,+3)$ Pits corresponding to local minima
of $\rho(\mathbf{r})$. The numbers of each type of CPs obbey the
following relationship depending on the nature of the system: 
$N(\mbox{peaks}) - N(\mbox{passes}) + N(\mbox{pales}) - N(\mbox{pits}) = 0 $
or $1$ 
for a crystal or an isolated system respectively. The Laplacian of the electron
density $\nabla^2 \rho(\mathbf{r})$ which is given by the trace of 
$H(\mathbf{r})$ is related to the
kinetic and potential electronic energy densities, respectively
G($\mathbf{r}$) and  V($\mathbf{r}$), by the local virial theorem,
$$ {1 \over 4} \nabla^2 \rho(\mathbf{r}) = 2 G(\mathbf{r}) + V(\mathbf{r})$$

\noindent
(atomic units are used throughout the paper). The sign of the
laplacian at a given point determines whether the positive kinetic energy or 
the negative potential energy density
is in excess. A negative (positive) Laplacian implies that density is locally
concentrated (depleted). Within the Quantum Theory of Atoms in Molecules, a 
basin is associated to each attractor $(3,-3)$ CP, defined as the region
containing all gradient paths terminating at the attractor. The boundaries of this
basin are never crossed by any gradient vector trajectory and satisfy
$\mathbf{\nabla}\rho(\mathbf{r}) \cdot \mathbf{N}(\mathbf{r}) = 0$, where 
$\mathbf{N}(\mathbf{r})$ is the normal to the surface at point $\mathbf{r}$. The 
corresponding surface is called  the zero flux surface and defines the atomic
basin when the attractor corresponds to a nucleus. Only in very rare cases 
non-nuclear attractors have been evidenced (Madsen et al., 1999). Within this
space partionning, the atomic charges deduced by integration over the
whole basin are uniquely defined.

During the last twenty years, several programs have been developped to perform
topology of electron densities but they are either 
connected to computer program packages 
(Gatti et al., 1994; Koritsanszky et al., 1995; Souhassou et al., 1999; 
Stash et al., 2001; 
Stewart et al. 1983; Volkov et al., 2000) or have limitations concerning
the type of wavefunctions which have been used to determine $\rho(\mathbf{r})$
in ab-initio calculations (Barzaghi, 2001; Biegler K\"onig et al., 1982; 
Biegler  K\"onig et al., 2001; Popelier, 1996 ) or to refine experimental data 
(Barzaghi, 2001). To our knowledge,only two cases were described to
analyse the topology of $\rho(\mathbf{r})$ numerically on grids;
Iversen et al.  (Iversen et al., 1995) used a maximum entropy density and 
Aray et al.
(Aray et al., 1997) sampled a theoretical density on a homogeneous
grid. However, these approaches are limited to the determination of
CPs.  The present analysis of the topological features of total
electron densities is independent of the way these densities 
are obtained and works for periodic or
non-periodic systems. We show in this paper that this can be simply achieved
by working with densities given on regular grids in real space. The
developed software InteGriTy uses a tricubic Lagrange interpolation
which makes the CP search and integration method both accurate and fast.
The densities used to illustrate the performance of our approach are
theoretical ab-initio densities obtained with the Projector Augmented Wave
(PAW) method (Bl\"ochl, 1994). The next section of this paper gives
a short description of the method. Test compounds and computational
details are given in Section 3.
Section 4 is devoted to the determination of CPs and their characteristics whereas
section 5 concerns the determination of atomic basins and charges.
We will discuss the effect of grid spacing of the input density and
the plane wave cutoff used for the PAW calculations on the properties of
different type of interactions (ionic, covalent and intermolecular).

\section{Description of the method}
\subsection{Input data and interpolation}

In order to achieve the topological analysis of any experimental or theoretical
electron density, the density is given on a regular,
not necessarily homogeneous grid in real space. A grid of stored
values of $\rho(\mathbf{r})$ must be prepared, preferably in binary format in order
to save disk space and with double precision to ensure high precision. The
grid which is defined by its origin, three meshgrid vectors and the number of points
in each grid direction as well as atomic positions must be specified with
respect to a cartesian coordinate system. Determinations of the topological 
properties of $\rho(\mathbf{r})$ requires the knowledge of $\rho(\mathbf{r})$,
$\mathbf{\nabla}\rho(\mathbf{r})$ and  $H(\mathbf{r})$ at many arbitrary points.
This can be achieved in an accurate and efficient way by using a tricubic 
Lagrange interpolation (Press et al., 1992). In one dimension, it uses values of $\rho$ on two
grid points on each side of the current point as illustrated in 
figure~\ref{lagrange}. For a three-dimensional system, it uses 64 grid points 
surrounding the box containing the current point:

$$ \rho(x,y,z) = \sum_{i=1}^{4}  \sum_{j=1}^{4}  \sum_{k=1}^{4}
\rho(i,j,k) \, L_i(x) \, L_j(y) \, L_k(z). $$

\noindent
As the first and second derivatives of this expression are straighforward,
the evaluation of $\mathbf{\nabla}\rho(\mathbf{r})$ and  $H(\mathbf{r})$ 
is numerically very cheap. 
It is clear that with respect to 
analytical expressions, the interpolation may introduce errors. However, as
shown in section 4 and 5 these errors are small 
for reasonable values of the grid interval size, 
and insignificant when compared to those 
issued from the multipolar refinement of experimental structure factors.
One should also notice that this interpolation is not
suited to perform the topology of $\nabla^2 \rho(\mathbf{r})$ from the
density. Higher order interpolation would be required or the Laplacian
has to be supplied on a grid.

\begin{figure}
\caption{One dimensional example for tricubic Lagrange interpolation.
$x_i$ and $\rho_i$ are respectively the abscissa and density value at grid
point $i$. The density at the current abscissa $x$ is given by $\rho(x)$
where $L_i(x)$ are third order polynomial, passing through all the grid points
as shows the example of $L_1(x)$. }
\begin{picture}(243,100)(-50,-80) 
\thicklines
\put(0,0){\vector(1,0){102}}
\multiput(7,0)(17,0){6}{\circle*{3.5}}
\put(21,7){$\rho_1$}
\put(38,7){$\rho_2$}
\put(55,7){$\rho_3$}
\put(72,7){$\rho_4$}
%\put(89,7){value of the density}
\put(21,-10){$x_1$}
\put(38,-10){$x_2$}
\put(55,-10){$x_3$}
\put(72,-10){$x_4$}
{\thinlines \put(50,-20){\vector(0,1){20}}}
\put(47,-29){$x$, $\displaystyle \rho(x) = \sum_{i=1}^4 L_i(x) \rho_i$}
\put(-12,-65){$\displaystyle L_1(x) = 
{(x-x_2)(x-x_3)(x-x_4) \over (x_1-x_2)(x_1-x_3)(x_1-x_4)}$}
\end{picture}
\label{lagrange}
\end{figure}

\subsection{Critical points}

To locate the CPs, starting from every grid point, 
a standard Newton-Raphson technique (Press et al., 1992)
is used to find the zero's of its gradient modulus:

$$ \mathbf{r}_{i+1} = \mathbf{r}_{i} - \alpha \; H^{-1}(\mathbf{r}_{i}) \cdot
\mathbf{\nabla}\rho(\mathbf{r}_{i}).$$

Far from the CPs, the full Newton step will not necessarily decrease
the gradient modulus and the parameter $\alpha$ allows the
stepsize adjustment. In all our calculations an  $\alpha$  value of $0.3$ led to stable
results. This iterative procedure is used until the gradient modulus
becomes less than a choosen threshold value. The corresponding CP is then
stored if no other CP has been found in its vinicity. Otherwise, the program
keeps the point which has the smallest gradient modulus. The CPs can then be 
classified with respect to their type and/or to the magnitude of 
$\rho(\mathbf{r}_{\mbox{\scriptsize CP}})$.
It is worth emphasizing that, since each grid point acts in its turn
as a starting  point, CP search does not require a priori
knowledge of atom location, nor the definition of plane or local
coordinates system. Periodic boundary conditions are used to treat periodic
systems whereas four grid points at each border of the input box
are ignored in the case of non periodic systems.

\subsection{Atomic Basins}

Interpolation of electronic density on grid can also be used to
derive atomic basins and integrate the density to obtain the
atomic charges with good accuracy
and reasonable computer time. The surface $S_{\Omega}$ of each basin
$\Omega$ is determined by its intersection with rays originating from
the attractor. Only one intersection per ray is looked for. Then the
determined surfaces may not be fully correct (Biegler K\"onig et al.,
1982; Popelier, 1998) but the missing volume
that can be checked a posteriori is very low, thus having no
appreciable effect on the integrated charges.
Basin search is performed on total density whereas highly accurate
integration is obtained from valence part only to avoid using
unreasonably small grid steps. 
The program works with both periodic and non periodic conditions for 
the input grids and a threshold electron density value can be applied
to limit the surface of open systems (e.g. van der Waals envelope, Bader, 1990). 
The present integration results concern only periodic systems thus
allowing a posteriori validation of the process according to the sum
over the whole unit cell of all basin volumes, including all atoms and 
possible non nuclear attractors.

\begin{figure}
\caption{Schematic diagram to illustrate how a running point in radial
  coordinates is checked following the gradient path inside $\left(
  \bullet \right)$  or outside $\left( \circ\right)$ a basin
  centered on the attractor A and delimited  by the surface
  $S_{\Omega}$. $\mathbf{r}_{A}$ and $\mathbf{r}$ respectively give
  the positions of the attractor and the running point in the absolute
  cartesian coordinates system. The incremental step length $g \, dr_o$ is
  defined in the text.}
\resizebox{9cm}{!}{
\includegraphics[20mm,215mm][110mm,280mm]{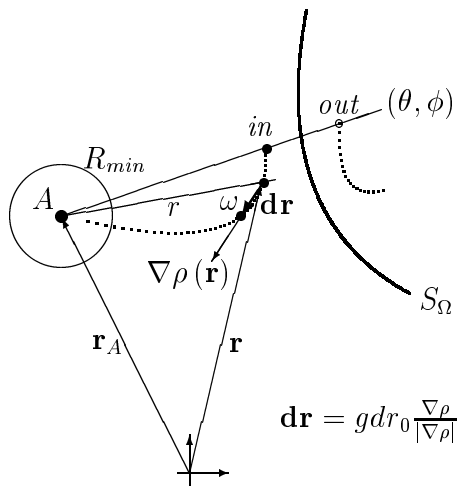}
}
\label{basin1}
\end{figure}

\subsubsection{Basin search}

\begin{figure}
\caption{Basin surface location flowchart for a given ray originating
  from an  attractor with spherical coordinates $\theta$ and
  $\phi$. $L$ and $H$ are dummy logical constants to check if both the
  \textit{low} and \textit{high} limits of the coarse bracketting have
  been found. In the second part of the flowchart, the dichotomy
  process is stopped when the distance between the two limits is below
  the predefined tolerance $d_{tol}$.}
\resizebox{9cm}{!}{
\includegraphics[55mm,50mm][175mm,280mm]{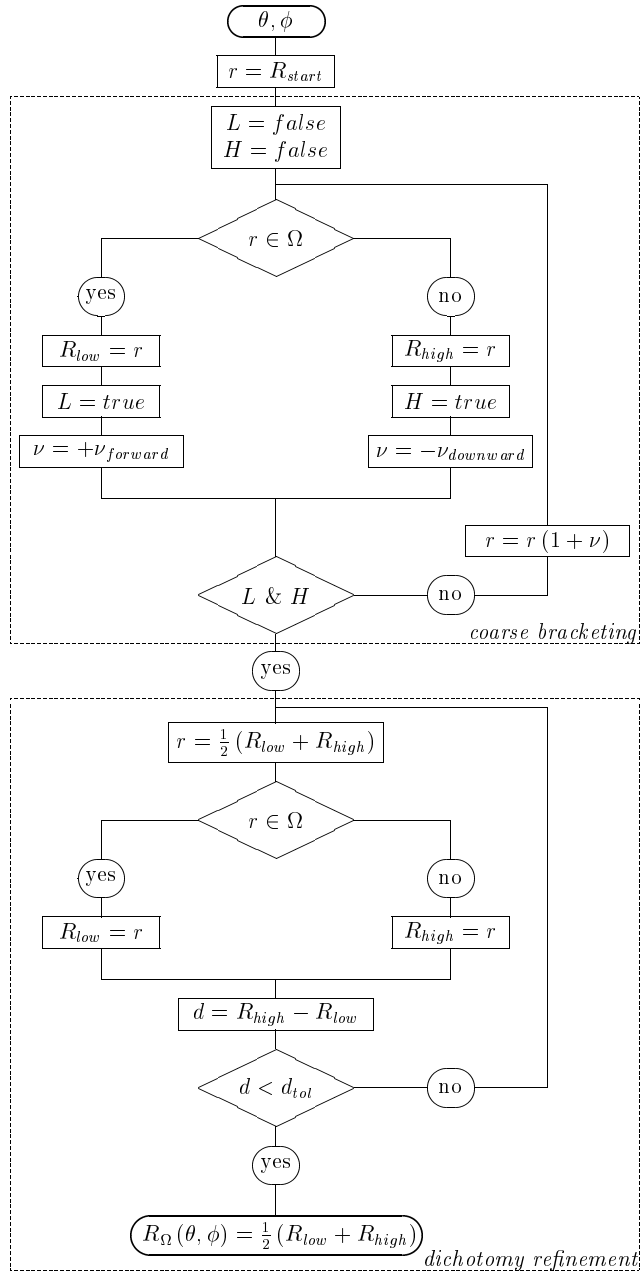}
}
\label{basin2}
\end{figure}

To search for the surface of a basin, a radial coordinates system
centered on each attractor is used and one point of the basin surface
  $R_{\Omega}\left( \theta, \phi \right)$
is looked for along the ray defined by $\theta$ and $\phi$ angles.
A point  $r\left( \theta, \phi \right)$ is declared inside the basin
$\Omega$ if the following gradient path brings it towards a sphere 
centered on the attractor and with radius $R_{min}$ small enough to 
be in the basin, as illustrated in figure~\ref{basin1}. 
The running point is declared outside of $\Omega$ if
a few iteration steps successively move it away from the attractor.
The step amplitude used to follow the gradient is the product of a
minimum step $dr_0$ weighted by a coefficient $g$ depending on
the angle $\omega$ between the considered ray and the gradient. It takes the form 
$ln\left(g\right)=A\mid cos\left(\omega\right)\mid ^B$ so that the
maximum step size corresponds to the parallel situation. 
 The search algorithm used for each ray is given in figure~\ref{basin2}. 
  A coarse bracketing, $R_{low} < R_{\Omega}\left( \theta, \phi
  \right) < R_{high}$,  is first performed starting from the value
 $R_{start}$ and using geometric progression with common ratio $1
+ \nu$.
The sign and amplitude of $\nu$ depends on wether
bracketing is done \textit{downward} to or \textit{away} from the attractor.
A dichotomy procedure is then used to refine the
$R_{\Omega}\left( \theta, \phi \right)$ value with predefined tolerance $d_{tol}$.
At the first $\left( \theta, \phi \right)$ step, $R_{start}$ is set to 
an arbitrary value given for each atom as an input parameter.  
A crude estimation of basins limits is done first on a regular $\left( \theta,
  \phi \right)$ grid with small number of points $n_{\theta}$ and
$n_{\phi}=2n_{\theta}$ with each $R_{\Omega}\left( \theta,\phi
\right)$ acting as starting value for the next $\left( \theta, \phi \right)$ step.
This search is sufficient for graphical purpose and enables $R_{start}$
initialisation by linear interpolation at all $\left( \theta,\phi
\right)$ points added during the integration process.  
The $R_{min}$ value is updated at the end of the crude estimation in order to save
time during integration process. It is automatically reset to a
lower value where it needs to be so.

\subsubsection{Integration}

Two methods can be used for integration, both using three nested
integration do loops with spherical coordinates. The first one is
straightforward and uses the same fixed number of
$\left(\theta,\phi,r \right)$ points for all attractors. The
number $n_{r}$ and  $n_{\theta}$ of radial and $\theta$ steps are kept
fixed, whereas the number $n_{\phi}\left(\theta\right)$ of $\phi$
steps is $\theta$ dependent such that elementary solid angle
$sin\left( \theta\right)\delta\theta\delta\phi$ be constant. The
integral $Q_{f}$ of a quantity $f\left(\theta,\phi\\,r \right)$ is
simply given by the discrete sum:
$$ Q_{f} = \delta r \, \delta\theta \, \sum_{i=1}^{n_{\theta}}
sin\left(\theta_{i}\right)  \delta \phi \left(\theta_{i}\right)
\sum_{j=1}^{n_{\phi}\left( \theta_{i} \right)} \sum_{k=1}^{n_{r}} r_{k}^{2} 
\, f \left( \theta_{i},\phi_{j},r_{k}\right) $$ 

\begin{figure}
\caption{Romberg flowchart for the integration of a function $f$ in
 the interval $[a,b]$. $Q_{f}$ stands for an integrated property (charge,
 laplacian,volume) at the three imbricated levels of integration,
  namely: radial, $\phi$ and $\theta$ in the respective intervals 
  $[0,R_{\Omega}\left(\theta,\phi\right)]$, $[0,2\pi]$ and $[0,\pi]$.}
\resizebox{9cm}{!}{
\includegraphics[55mm,123mm][175mm,280mm]{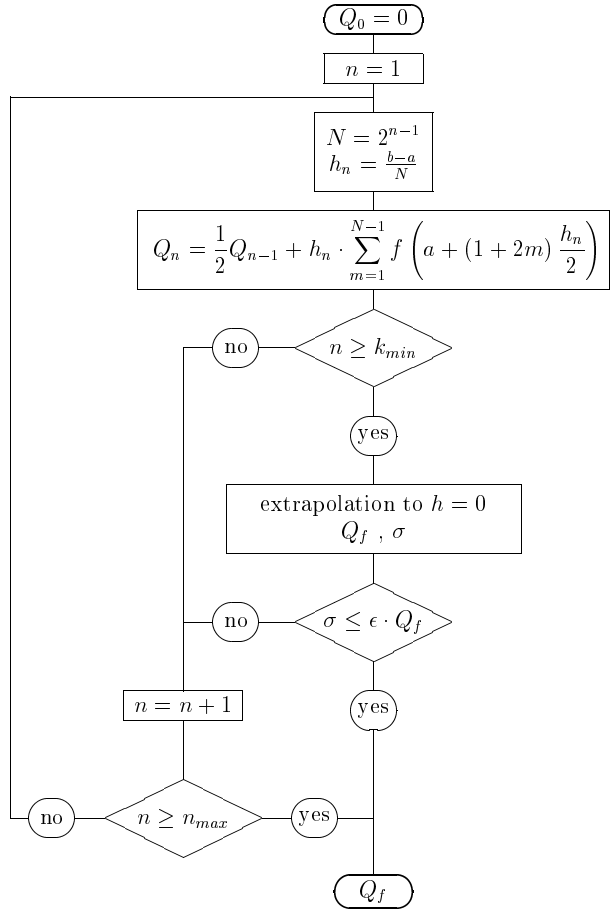}}
\label{romberg1}
\end{figure}

The second method uses Romberg procedure (Press et al., 1992) and is 
illustrated in figure~\ref{romberg1} for a single variable function
$f$  to be integrated in the interval $[a,b]$.
It first calculates an estimation of the integral $Q_{f}$ over $n$
stages  of the extended trapezoidal rule with successive integration 
steps $\{h_n\}$ and extrapolates $Q_{f}$ with  $k$ order polynomial 
in $h^{2}$ to the continuum limit $h=0$ over the last $n-k$
stages. The interpolation starts after n$ \geq k_{min}$ stages.
This iterative process is terminated when either an estimated error
$\sigma$ derived from  the extrapolation operation yields a desired 
relative accuracy $\epsilon$ such that $\sigma \leq (\epsilon \cdot
Q_{f})$ or a fixed maximum number of stages $n_{max}$ is reached.
Several convergence criteria can be used for the radial, $\phi$ and
$\theta$ levels of integration. Total and valence atomic charges,
volume and laplacian are integrated and convergence criteria
can be choosen on each of these quantities. Individual quantities are 
summed up at the end of the loop over all attractors. Total charges
and volume are then compared to the expected ones and residuals give 
the accuracy of the integration process.\\

\section{Test compounds}

Different types of crystals and molecules have been used to test limits, 
accuracy and performance of algorithm. In this paper, we have 
selected two crystals, NaCl a classical example for
ionic crystals, and TTF-2,5Cl$_2$BQ for covalent and intermolecular
interactions. The latter compound was also chosen for the following reasons: 
with 26 atoms in the unit cell
the system is neither too small nor too large; the unit cell is 
triclinic then the grid will not be orthogonal; the presence of inversion 
symmetry should be recovered in all properties; finally 
the expected small intermolecular charge transfer 
(about 0.5 out of 200 electrons) is a good
quantity to test for the accuracy of charge integration.
\begin{figure}
\caption{Representation of a mixed stack of alternating TTF and 2,5Cl$_2$BQ
molecules including atomic numbering. Open circles indicate bond and ring
CPs. The lines connecting two molecules correspond to  
bond paths determined by following
$\mathbf{\nabla}\rho(\mathbf{r})$ with a small stepsize.
Diamonds indicate some intermolecular $(3,-1)$ CPs.}
\centerline{\resizebox{8.0cm}{!}{
\includegraphics{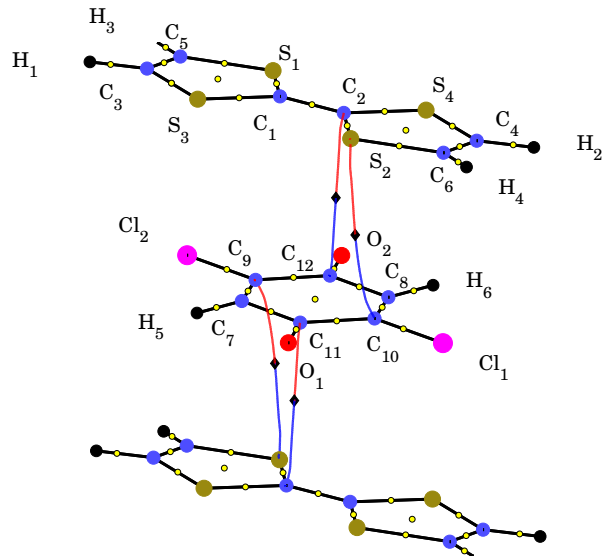}}}
\label{chainTTFBQ}
\end{figure}
All calculations used to generate the electron densities were carried
out with the Projector Augmented Wave (PAW) method (Bl\"ochl, 1994), 
an all-electron code developped by P.~E.~Bl\"ochl. The wave functions
were expanded into augmented plane wave up to a kinetic energy 
(E$_{\mbox{\scriptsize cutoff}}$) ranging
from 30 to 120 Ry. NaCl was treated within an fcc cell with a = 10.62 a.u.
and 8 $\mathbf{k}$ points in reciprocal space. For TTF-2,5Cl$_2$BQ we used the
experimental geometry at ambient conditions 
(Girlando et al., 1993). The unit cell is triclinic (a = 14.995, b = 13.636, 
c = 12.933 a.u., $\alpha$ = 106.94$^\circ$, $\beta$ = 97.58$^\circ$ 
and $\gamma$ = 93.66$^\circ$) and contains one TTF
and one 2,5Cl$_2$BQ molecule both setting on inversion centers. These molecules
are respectively electron donor and acceptor molecules alternating along the
\textbf{b} axis to form mixed stacks (figure~\ref{chainTTFBQ}). 
The ab-initio calculations were performed with three $\mathbf{k}$
points between $\Gamma$ and $Y={1 \over 2} \mathbf{b}^{*}$ in the
Brillouin Zone (Katan et al., 1999).

%-------------------------------------
\section{Critical points}

\subsection{Ionic bonds} 
\begin{figure}
\caption{Octahedron with one Cl at its center and six Na at each vertex. Iso-density
curves represented in the plane of Na$\cdots$Cl nearest neighbors. The six dark spheres
correpond to the $(3,-1)$ CPs between Na and Cl, the twelve grey spheres to the
$(3,-1)$ CPs between Cl$\cdots$Cl and the bright smallest one to the $(3,+1)$ CPs.
The $(3,+3)$ CPs are not shown here.}
\includegraphics{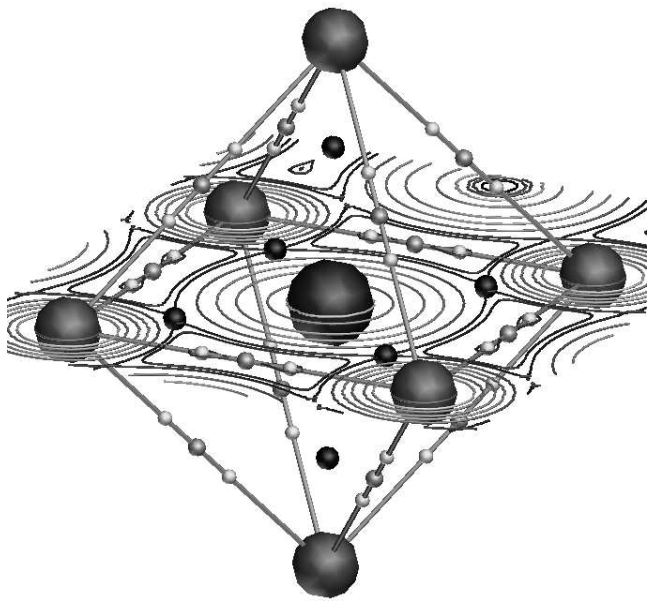}
\label{NaCl-cp}
\end{figure}

NaCl presents
four types of different CPs as shown in figure~\ref{NaCl-cp}. Within the unit cell
there are six $(3,-1)$ CPs between Na and Cl, six $(3,-1)$ CPs between Cl$\cdots$Cl 
surrounded in the Na$\cdots$Na direction
by twelve $(3,+1)$ CPs, two $(3,+3)$ CPs and two nuclear
attractors. The characteristics of these CPs are given in table 1
along with the variation of CP  properties versus grid spacing ($\Delta
r_{\mbox{\scriptsize grid}}$) and number of plane waves
(E$_{\mbox{\scriptsize cutoff}}$) for only one representative CP, all
other CPs presenting even smaller variations.  These results show that 
a grid spacing of about 0.15 a.u. and a plane wave cutoff of 30 Ry are 
sufficient to achieve accurate results for this ionic compound.
\subsection{Molecular compounds}
\begin{figure}
\caption{Representation of atomic basins in the plane containing both
  TTF and 2,5Cl$_2$BQ molecules. The strongest hydrogen bond in
  TTF-2,5Cl$_2$BQ crystal occurs between O$_1$ and H$_3$. The arrows
  indicate the direction and magnitude of the gradient in the
  plane. An example of unreachable small piece of volume due to
  multiple intersection of the atomic surface with a ray originating
  from atom C$_5$ can be seen on top right of the figure.}
\includegraphics{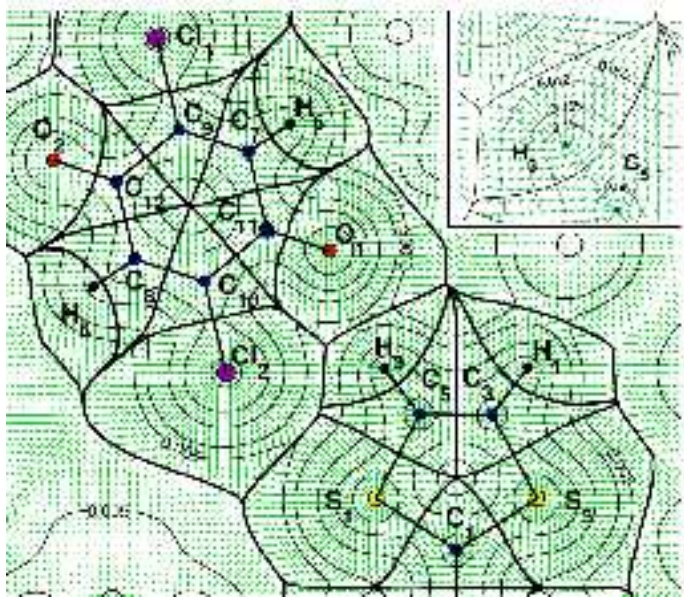}
\label{plane-basin}
\end{figure}

Low densities at critical points are observed in the crystal TTF-2,5Cl$_2$BQ
in interaction regions.  Two examples are selected on table 2 and 3
respectively corresponding to the ring
CP of  2,5Cl$_2$BQ and to the strongest hydrogen bond occuring
in the plane shown in figure~\ref{plane-basin}. In both cases, the smallest
E$_{\mbox{\scriptsize cutoff}}$ (50 Ry) and the largest grid spacing (0.125 a.u.)
give already quite good results. These properties converge even better for
all other low density CPs. As for ionic bonding, the electron density is
particularly smooth close to the CPs and even smaller 
E$_{\mbox{\scriptsize cutoff}}$ and
$\Delta r_{\mbox{\scriptsize grid}}$ are enough.
The situation is much different for covalent bonds where 
$\rho(\mathbf{r}_{\mbox{\scriptsize CP}})$ is much higher and the
electron density varies more rapidly. All covalent bond CPs have
been plotted on figure~\ref{chainTTFBQ}. Typical covalent bonds of
TTF-2,5Cl$_2$BQ are summarized in table 4 and 5 for different  
E$_{\mbox{\scriptsize cutoff}}$ and $\Delta r_{\mbox{\scriptsize grid}}$. 
For most bonds, the CP properties do not vary too much, except
for C=O bond where low E$_{\mbox{\scriptsize cutoff}}$ gives a
value of $\nabla^2 \rho(\mathbf{r}_{\mbox{\scriptsize CP}})$ 
twice as large as other E$_{\mbox{\scriptsize cutoff}}$.
This is due to the electron density that 
changes very rapidly along the bond path. For this reason, the  C=O bond
properties are also the most sensitive to $\Delta r_{\mbox{\scriptsize grid}}$
(table 5). This makes it particularly difficult to treat. Evidently, 
the CP characteristics of simple bonds are more easily accurate than those 
of double bonds.

Finally, we have checked in each case that CPs which are equivalent by
symmetry are equivalent at least within
the numbers of digits indicated in the different tables of this section. All
these results clearly show that CPs properties can easily be deduced from
densities given on regular grids, the choice of the grid stepsize
depending on the nature of the bonds of interest.

\section{Atomic basins}

Accuracy of integration from a grid density will depend on the integration
method and on the way of determining the basins surfaces but also on the
accuracy of electron density data at the grid points, on the degree of
missing information due to grid spacing and on the possible bias
introduced by the interpolation procedure. The following residuals
have been defined to estimate the accuracy of integration process:
N$_{\mbox{\scriptsize grid}}$ is the number of electrons in the unit
cell obtained over all elementary volume units either by
discrete summation  or using analytical integral expression 
of the tricubic interpolation.
Then $\Delta$N = N$_{\mbox{\scriptsize grid}}$-N$_{\mbox{\scriptsize real}}$ is
the difference between the number of valence electrons obtained by
integration over the whole unit cell and its expected real value. The
quantities v$_\Omega$ and S$_\Omega$ respectively refer to  the volume and surface of
the basin $\Omega$. The number of electrons in an
individual basin is denoted  n$_\Omega$.
 $\delta$N = $\sum \mbox{\scriptsize n}_{\Omega}$ -
 N$_{\mbox{\scriptsize real}}$ is the residual after summation over all atomic basins.
The basin volume uncertainty is set equal to the product of the tolerance
d$_{tol}$ and an estimation of the atomic surface S$_\Omega$. The
total volume uncertainty is given by $\sigma_{\mbox{\scriptsize V}}$ = $d_{tol} \, \sum
\mbox{\scriptsize S}_{\Omega}$ and the residual volume error by $\Delta$V
=  $\sum \mbox{\scriptsize v}_{\Omega}$ - V$_{\mbox{\scriptsize
    cell}}$.
If the uncertainty $\sigma_{\mbox{\scriptsize V}}$ is found clearly less than the residual
$\Delta$V, the question arises about the way the atomic
surfaces are derived. In this case, either there are several intersections of the
basin surface with one ray originating from the attractor,
either the parameters used to follow the gradient path have to be
modified or some attractors are missing in the input list. 
Finally the largest charge difference between symmetry equivalent
atoms, the intra or inter molecular charge transfer (CT) and its estimated uncertainty
($\sigma_{\mbox{\scriptsize CT}}$) are the other criteria which can be
used as a measure of the accuracy of the integration process.

\subsection{Grid spacing and \textit{ab-initio} calculation convergence}

\begin{figure}
\caption{NaCl atomic charges from total electron density vs. grid
  spacing. Solid lines are guides for the eyes converging to the 0.83
  charge transfer estimated from valence density.}
\centerline{\resizebox{7.5cm}{!}{\includegraphics{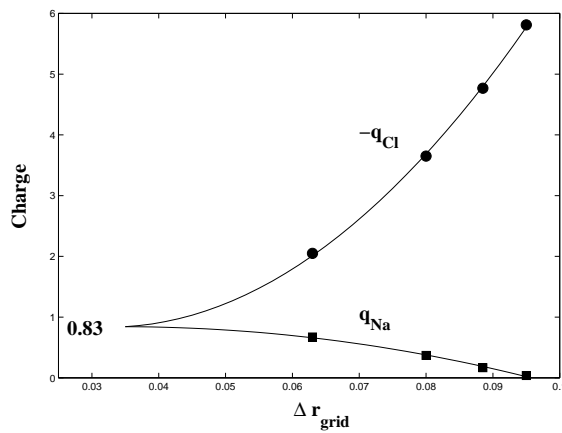}}}
\label{charges1}
\end{figure}

  In the case of NaCl, \textit{ab-initio} calculations are not too
  sensitive to the convergence criterion E$_{\mbox{\scriptsize cutoff}}$
  so the grid spacing effect can clearly be evidenced. Well converged integration
  results using Romberg procedure are given in
  table~\ref{tableI1}. One can notice that the electron number
  residual  $\delta$N after integration
  and sum  over all basins is very close to $\Delta$N given as
  input. This residue monotonically decreases with
  reducing grid spacing whereas volume residual  $\Delta$V is almost
  constant.  Reasonnable estimation of interatomic charge transfer can
  be derived from valence density within a precision of 0.01 electron
  for a grid spacing of about 0.06 a.u. The prohibitive
  grid spacing required to get the charge transfer 
  with the same precision from  total density can be estimated to 0.03
  a.u. as illustrated in figure~\ref{charges1}. 
 \subsection{Grid spacing and \textit{ab-initio} calculations convergence}
\begin{figure}
\caption{3D representation of some atomic basins in the TTF-2,5Cl$_2$BQ charge
 transfer complex.}
\centerline{\resizebox{8cm}{!}{
\includegraphics{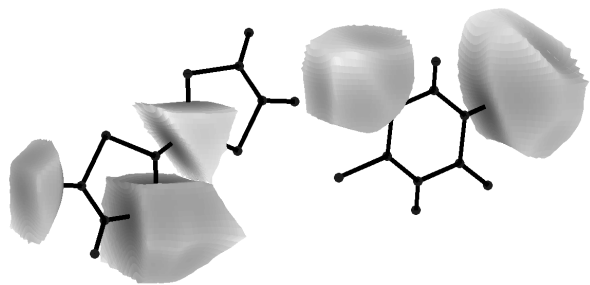}
}}
\label{3Dbasins}
\end{figure}

 In the case of molecular crystals which exhibit short bonds with large
 and quickly varying electron density at their bond critical points,
 such C=O for example, the \textit{ab-initio} calculations  are more
 sensitive to the convergence criterion E$_{\mbox{\scriptsize
 cutoff}}$ defining the plane waves expansion basis set.
 This is found in the integration results as illustrated in
 table~\ref{tableI2} for TTF-2,5Cl$_2$BQ compound where the electron number 
 residual $\delta$N after integration over all atomic basins is
 sligthly different from that after direct summation over the unit cell $\Delta$N. 
 This may arise from the larger  number of atoms and from the less
 smooth shapes of atomic basins (figure~\ref{3Dbasins}) in comparison
 to the NaCl case. Nevertheless using E$_{\mbox{\scriptsize cutoff}}$
 above 50Ry is enough to get the intermolecular charge transfer within
 0.02 electron precision which is the goal that originally motivated this work.  
\subsection{\textit{Romberg} vs. \textit{fixed spherical grid} integration }
Different sets of parameters for the \textit{Romberg} integration
procedure, including adequate parameters for the basin limits search are
given in table~\ref{tableI3}. Indicated CPU time corresponds to a PC
machine with 800MHz processor and 512Mb RAM without including
density files reading nor direct unit cell summation. The
preliminary estimation of basin envelopes is done with $n_\theta$ = 18 and
$n_\phi$ = 36, using a tolerance $d_{tol}$ = 0.05 a.u. This operation
takes about 0.8 second per basin. The integration convergence criteria have
been put only on the volume and the valence density since the grid
spacing used may not correctly restore density cusps near atomic nuclei.
Obtaining the intermolecular charge transfer, in the case of
TTF-2,5Cl$_2$BQ, within a precision of about one hundredth of an electron
takes about one minute per basin. The weaker criteria, although
leading to unsatisfying residual from the intermolecular charge
transfer point of view, nevertheless give a good indication of atomic
charges within less than one second per basin.
The residual volume error remains of the order of one cubic atomic
unit out of nearly 2500 for the unit cell and only for the more severe
integration convergence criteria may appear the problem of very
precise determination of basin boundaries
(figure~\ref{plane-basin}, top right). The greatest discrepancy between
charges obtained for equivalent atoms ranges from $10^{-4}$ for the set of
parameters leading to the best integration, to $10^{-3}$ for the set
giving the shorter CPU time. 
For comparison, integration has been done using a \textit{fixed spherical grid}
 for all basins. The corresponding parameters and results are listed in table~\ref{tableI4}.
The total number of angular loops is denoted $n_{\theta,\phi}$. Below
 $n_{\theta,\phi} \times n_r \sim 10^4$ integration
points, no realistic atomic charges can be obtained. 
Indeed, the very short time that can be used with \textit{Romberg} 
integration come from its internal $k$ order polynomial 
extrapolation which gives an error estimate of order
$O(1/N^{2k})$  instead of $O(1/N^{2})$ for the simple discrete 
summation with the same number $N$ of integration points.

In the case of  \textit{fixed spherical grid} integration, compared to
\textit{Romberg} integration, the volume uncertainty is in most of the
cases greater than the residual volume error, thus giving no
information about missing or overlapping volumes. The discrepancy
between charges of equivalent atoms is also about an order of
magnitude higher. The point that makes the \textit{Romberg} procedure more
favorable is that the number of integration points is
automatically adapted to each basin to the desired level of
convergence, as illustrated for atoms C$_1$ and S$_1$  in
table~\ref{tableI3}, whereas taking a fixed mean spherical grid leads to
missing or biased information for some atoms and adds unnecessary
points for other ones. The numbers $\overline{\mbox{\scriptsize
    n}}_{r}$ of radial loops indicated in
table~\ref{tableI3} are average over all angular loops for each
basin.  

\section{Conclusion}

We have shown that topological properties can be derived from electron
densities given on 3D grids with tricubic interpolation to
extract at any given position the density, its gradient and hessian
matrix. Except for very short covalent bonds such C=0, critical points
properties can be obtained with grid spacing of about 0.1 a.u. The
same  grid spacing can also be used to get atomic charges by
integration over atomic basins with highly accurate values using the
valence electron density. The integration is done with spherical
coordinates system centered on the attractors and uses the robust
\textit{Romberg} algorithm which allows the number of integration
points to be automatically adapted for each basin and offers the
choice between saving computing time or giving the preference to accuracy.

     %-------------------------------------------------------------------------
     % The back matter of the paper - acknowledgements and references
     %-------------------------------------------------------------------------

     % Acknowledgements come after the appendices

\ack{Acknowledgements}

This work has benefited from collaborations within the
$\Psi_k$-ESF Research Program and the Training and Mobility
of Researchers Program ``Electronic Structure'' (Contract: FMRX-CT98-0178)
of the European Union and the Interational Joint Research Grant
``Development of charge transfert materials with nanostructures''
(Contract: 00MB4). Parts of the calculations have been supported
by the ``Centre Informatique National de l'Enseignement Sup\'erieur''
(CINES---France). We would like to thank P.E. Bl\"ochl for his PAW
code and P. Sablonniere for usefull discussions.

     % References are at the end of the document, between \begin{references}
     % and \end{references} tags. Each reference is in a \reference entry.

\begin{table}
\caption{Characteristics of crystalline NaCl critical points (CPs) for different
grid spacing ($\Delta r_{\mbox{\scriptsize grid}}$) and numbers of plane waves
 (E$_{\mbox{\scriptsize cutoff}}$). $\lambda_1$, $\lambda_2$ and $\lambda_3$ are
the  Hessian matrix  eigenvalues. All values except  E$_{\mbox{\scriptsize cutoff}}$
are given in a.u.}
\begin{tabular}{lrcccc}
\hline
 Type & E$_{\mbox{\scriptsize cutoff}}$ & $\Delta r_{\mbox{\scriptsize grid}}$ 
& $\rho(\mathbf{r}_{\mbox{\scriptsize CP}})$  
& $\nabla^2 \rho(\mathbf{r}_{\mbox{\scriptsize CP}})$ 
& $\lambda_1$, $\lambda_2$, $\lambda_3$ \\
\hline
     (3,-1)& 40 Ry   & 0.1564 & 0.0130  & 0.0577  &-0.0122, -0.0119, 0.0817   \\
     NaCl  & 40 Ry   & 0.1138 & 0.0130  & 0.0576  &-0.0122, -0.0121, 0.0820   \\
           & 40 Ry   & 0.0939 & 0.0130  & 0.0571  &-0.0122, -0.0121, 0.0814   \\
           & 30 Ry   & 0.1138 & 0.0130  & 0.0581  &-0.0123, -0.0121, 0.0825   \\
           &120 Ry   & 0.1138 & 0.0130  & 0.0583  &-0.0123, -0.0121, 0.0827   \\
\hline
     (3,-1)& 40 Ry   & 0.1138 & 0.0049  & 0.0127  &-0.0024, -0.0011, 0.0163   \\
     ClCl \\
\hline
     (3,+1)& 40 Ry   & 0.1138 & 0.0047  & 0.0134  &-0.0025,  0.0035, 0.0124   \\
\hline
     (3,+3)& 40 Ry   & 0.1138 & 0.0019  & 0.0056  & 0.0019,  0.0019, 0.0019   \\
\end{tabular}
\end{table}

\begin{table}
\caption{Characteristics of the ring CP of 2,5Cl$_2$BQ and the O$_1 \cdots$H$_3$
hydrogen bond of TTF-2,5Cl$_2$BQ for different 
E$_{\mbox{\scriptsize cutoff}}$ and $\Delta r_{\mbox{\scriptsize grid}} = 0.104$ a.u.}
\begin{tabular}{lrccccc}
\hline
 Type &  E$_{\mbox{\scriptsize cutoff}}$ & $\rho(\mathbf{r}_{\mbox{\scriptsize CP}})$
  &$\nabla^2 \rho(\mathbf{r}_{\mbox{\scriptsize CP}})$ & $\lambda_1$ &
  $\lambda_2$ & $\lambda_3$ \\
\hline
Ring CP& 50 Ry   & 0.0210  & 0.1233  & -0.0128  & 0.0613   & 0.0747   \\
2,5Cl$_2$BQ& 70 Ry   & 0.0209  & 0.1327  & -0.0134  & 0.0664   & 0.0796   \\
       & 90 Ry   & 0.0209  & 0.1314  & -0.0131  & 0.0656   & 0.0789   \\
       &100 Ry   & 0.0209  & 0.1315  & -0.0131  & 0.0656   & 0.0791   \\
\hline
O$_1 \cdots$H$_3$& 50 Ry   & 0.0103  & 0.0359  & -0.0106  & -0.0102  & 0.0567   \\
          & 70 Ry   & 0.0103  & 0.0384  & -0.0109  & -0.0103  & 0.0596   \\
          & 90 Ry   & 0.0103  & 0.0375  & -0.0108  & -0.0103  & 0.0585   \\
          &100 Ry   & 0.0103  & 0.0370  & -0.0107  & -0.0103  & 0.0580   \\
\end{tabular}
\end{table}
\begin{table}
\caption{Same as table 2 for different $\Delta r_{\mbox{\scriptsize grid}}$ and
E$_{\mbox{\scriptsize cutoff}}= 90$ Ry.}
\begin{tabular}{lcccccc}
\hline
 Type &  $\Delta r_{\mbox{\scriptsize grid}}$  & $\rho(\mathbf{r}_{\mbox{\scriptsize CP}})$
  &$\nabla^2 \rho(\mathbf{r}_{\mbox{\scriptsize CP}})$ 
                               & $\lambda_1$ & $\lambda_2$ & $\lambda_3$ \\
\hline
Cycle  & 0.125  & 0.0209  & 0.1314  & -0.0132  & 0.0656   & 0.0790   \\
       & 0.104  & 0.0209  & 0.1314  & -0.0131  & 0.0656   & 0.0789   \\
       & 0.085  & 0.0209  & 0.1313  & -0.0131  & 0.0656   & 0.0788   \\
\hline
\hline
O$_1 \cdots$H$_3$& 0.125  & 0.0103  & 0.0375  & -0.0108  & -0.0102  & 0.0585   \\
          & 0.104  & 0.0103  & 0.0375  & -0.0108  & -0.0103  & 0.0585   \\
          & 0.085  & 0.0103  & 0.0373  & -0.0107  & -0.0103  & 0.0584   \\
\end{tabular}
\end{table}
\begin{table}
\caption{Characterisitcs of selected covalent bond CPs of TTF-2,5Cl$_2$BQ for
various E$_{\mbox{\scriptsize cutoff}}$; $\Delta r_{\mbox{\scriptsize grid}}$ = 0.104 a.u.}
\begin{tabular}{lrccccc}
\hline
 Type &  E$_{\mbox{\scriptsize cutoff}}$ & $\rho(\mathbf{r}_{\mbox{\scriptsize CP}})$
  &$\nabla^2 \rho(\mathbf{r}_{\mbox{\scriptsize CP}})$ 
                               & $\lambda_1$ & $\lambda_2$ & $\lambda_3$ \\
\hline
C$_{11}$-O$_1$& 50 Ry   & 0.4038  & 0.6240  & -1.0511  & -0.9774  &  2.6526  \\
          & 70 Ry   & 0.4070  & 0.3862  & -1.0359  & -0.9567  &  2.3788  \\
          & 90 Ry   & 0.4079  & 0.3182  & -1.0401  & -0.9579  &  2.3162  \\
          &100 Ry   & 0.4079  & 0.3131  & -1.0394  & -0.9571  &  2.3096  \\
\hline
C$_6$-S$_2$& 50 Ry   & 0.2089  & -0.3931 & -0.3364  & -0.2823  &  0.2256  \\
          & 70 Ry   & 0.2094  & -0.4185 & -0.3370  & -0.2835  &  0.2021  \\
          & 90 Ry   & 0.2094  & -0.4243 & -0.3371  & -0.2833  &  0.1961  \\
          &100 Ry   & 0.2094  & -0.4251 & -0.3371  & -0.2833  &  0.1954  \\
\hline
C$_1$-C$_2$& 50 Ry   & 0.3410  & -1.2044 & -0.7568  & -0.5956  &  0.1480  \\
          & 70 Ry   & 0.3387  & -1.0740 & -0.7496  & -0.5893  &  0.2649  \\
          & 90 Ry   & 0.3388  & -1.0786 & -0.7453  & -0.5853  &  0.2519  \\
          &100 Ry   & 0.3387  & -1.0734 & -0.7452  & -0.5852  &  0.2570  \\
\end{tabular}
\end{table}
\begin{table}
\caption{Same as table 4 for different $\Delta r_{\mbox{\scriptsize grid}}$;
 E$_{\mbox{\scriptsize cutoff}}$=90 Ry.}
\begin{tabular}{lcccccc}
\hline
 Type & $\Delta r_{\mbox{\scriptsize grid}}$  & $\rho(\mathbf{r}_{\mbox{\scriptsize CP}})$
  &$\nabla^2 \rho(\mathbf{r}_{\mbox{\scriptsize CP}})$ 
                               & $\lambda_1$ & $\lambda_2$ & $\lambda_3$ \\
\hline
C$_{11}$-O$_1$& 0.125  & 0.4080  & 0.2268  & -1.0899  & -1.0180  &  2.3347  \\
          & 0.104  & 0.4079  & 0.3182  & -1.0401  & -0.9579  &  2.3162  \\
          & 0.085  & 0.4079  & 0.2913  & -1.0266  & -0.9879  &  2.3058  \\
\hline
C$_6$-S$_2$& 0.125  & 0.2095  & -0.4207 & -0.3353  & -0.2824  &  0.1970  \\
          & 0.104  & 0.2094  & -0.4243 & -0.3371  & -0.2833  &  0.1961  \\
          & 0.085  & 0.2095  & -0.4239 & -0.3369  & -0.2835  &  0.1965  \\
\hline
C$_1$-C$_2$& 0.125  & 0.3388  & -1.0770 & -0.7438  & -0.5841  &  0.2509  \\
          & 0.104  & 0.3388  & -1.0786 & -0.7453  & -0.5853  &  0.2519  \\
          & 0.085  & 0.3388  & -1.0804 & -0.7462  & -0.5862  &  0.2520  \\
\end{tabular}
\end{table}
\begin{table}
\caption{NaCl integration results vs.  $\Delta r_{\mbox{\scriptsize grid}}$;
 E$_{\mbox{\scriptsize cutoff}}$=120 Ry.}
\begin{tabular}{lrrrr}
\hline
 $\Delta r_{\mbox{\scriptsize grid}}$ & 0.095 & 0.0885 & 0.08 & 0.063\\
\hline
$\Delta$N  & 0.0362 & 0.0289 & 0.0198 & 0.0086 \\
\hline
$\delta$N  & 0.0367 & 0.0285 & 0.0199 & 0.0090 \\
\hline
$\Delta$V  & 0.095 & 0.096 & 0.020 & 0.090    \\
\hline
q$_{\mbox{\scriptsize Na}}$ & 0.8282 & 0.8282 & 0.8188 & 0.8283      \\
v$_{\mbox{\scriptsize Na}}$ & 63.36 & 63.36 & 63.29 & 63.36 \\
\hline
q$_{\mbox{\scriptsize Cl}}$ &-0.8648 &-0.8572 &-0.8387 &-0.8373 \\
v$_{\mbox{\scriptsize Cl}}$ & 236.18 & 236.18 & 236.17 & 236.17 \\
\hline
CT & 0.85      & 0.84      & 0.83      & 0.83     \\ 
$\sigma_{\mbox{\scriptsize CT}}$   & 0.04 & 0.03 & 0.02 & 0.01 
\end{tabular}
\label{tableI1}
\end{table}

\begin{table}
\caption{TTF-2,5Cl$_2$BQ integration results vs. E$_{\mbox{\scriptsize
      cutoff}}$ ( $\Delta r_{\mbox{\scriptsize grid}}$ = 0.104 a.u.);.}
\begin{tabular}{lrrrr}
\hline
E$_{\mbox{\scriptsize cutoff}}$ (Ry) & 50 & 70 & 90 & 100\\
\hline
$\Delta$N     &-0.006 &-0.005 &-0.006 &-0.006 \\
\hline
$\delta$N     & 0.066 & 0.022 & 0.020 & 0.020 \\
\hline
$\Delta$V     &-0.92 &-1.44 &-1.93 &-1.89 \\
\hline
q$_{\mbox{\scriptsize TTF}}$         & 0.455 & 0.453 & 0.452 & 0.453 \\
q$_{\mbox{\scriptsize 2,5Cl$_2$BQ}}$ &-0.522 &-0.475 &-0.472 &-0.473 \\
\hline
CT      & 0.49 & 0.46 & 0.46 & 0.46   \\
$\sigma_{\mbox{\scriptsize CT}}$  & 0.07 & 0.02 & 0.02 & 0.02 
\end{tabular}
\label{tableI2}
\end{table}

\begin{table}
\caption{Sets of tested parameters and results for \textit{Romberg}
  integration on the charge transfer crystal TTF-2,5Cl$_2$BQ. The
  cell volume is V$_{\mbox{\scriptsize cell}}$ = 2493.062 a.u. 
  Direct integration of valence electron density over the whole unit
  cell gives an input error $\Delta$ N  = -0.0056 electron.
  $\Delta r_{\mbox{\scriptsize grid}}$ = 0.104 a.u; 
  E$_{\mbox{\scriptsize cutoff}}$ = 90 Ry }

\begin{tabular}{lrrrrrr}
\hline
\textit{Basin search} & & & & & & \\
$dr_o = d_{tol}$ & 5 10$^{-2}$ & 5 10$^{-3}$ & 10$^{-3}$  & 5
10$^{-4}$  &  5 10$^{-4}$  & 10$^{-4}$  \\
 A & 5 & 50 & 250 & 500 & 500 & 2500 \\
 B & 0.5 & 0.5 & 0.5 & 1 & 1 & 1 \\
\hline
\textit{Romberg} & & & & & & \\
$k_{min}$ & 4 & 6 & 6 & 6 & 6 & 6 \\
$\epsilon_{r}$ & 10$^{-3}$ & 10$^{-3}$ & 5 10$^{-4}$ & 10$^{-4}$ &
2 10$^{-7}$ & 10$^{-7}$ \\
$\epsilon_{\phi}$    & 10$^{-2}$ & 10$^{-2}$ & 5 10$^{-3}$ & 10$^{-3}$
& 8 10$^{-5}$ & 10$^{-5}$ \\
$\epsilon_{\theta}$  & 5 10$^{-2}$ & 5 10$^{-2}$ & 10$^{-2}$ &
10$^{-3}$ & 5 10$^{-5}$ & 10$^{-4}$ \\
\hline
\textit{CPU} time & 23s & 104s & 222s & 19min. & 3.5h & 43h \\
\hline
$\delta$N   &-0.413 & 0.080 & 0.056 & 0.019 & 0.014 & 0.006 \\
\hline
$\Delta$V   & 18.98 &-1.88 &-0.98 & -1.56 &-0.77 &-1.11 \\
$\sigma_{\mbox{\scriptsize V}}$   & 120 & 12 & 2.4 & 1.25 & 1.2 & 0.2\\
\hline
q$_{\mbox{\scriptsize TTF}}$ & 0.862 & 0.424 & 0.434 & 0.448 & 0.454 & 0.455 \\
q$_{\mbox{\scriptsize 2,5Cl$_2$BQ}}$ &-0.448 &-0.503 &-0.490 &-0.467 &-0.468 &-0.461 \\
\hline 
CT    & 0.65 & 0.46  & 0.46 & 0.46 & 0.461 & 0.458 \\
$\sigma_{\mbox{\scriptsize CT}}$ & 0.40 & 0.08  & 0.06 & 0.02 & 0.015 & 0.006 \\
\hline
\textbf{C$_1$} & & & & & &  \\
$n_ {\theta,\phi}$ & 89 & 1089 & 1089 & 1441 & 8639 & 38081 \\
$\overline{\mbox{\scriptsize n}}_{r}$ & 15 & 33 & 33 & 56 & 212 & 258 \\
q         &-0.311 &-0.285 &-0.285 &-0.283 &-0.284 &-0.284 \\
v         & 69.11 & 65.96 & 65.96 & 65.50 & 65.56 & 65.56 \\
\hline
\textbf{S$_1$} & & & & & &  \\
$n_ {\theta,\phi}$  & 81 & 1089 & 1089 & 1089 & 2591 & 8513 \\
$\overline{\mbox{\scriptsize n}}_{r}$ & 23 & 33 & 40 & 61 & 292 & 373 \\
q         & 0.249 & 0.252 & 0.255 & 0.258 & 0.258 & 0.258 \\
v         & 183.22 & 181.86 & 181.84 & 181.95 & 182.21 & 182.18 \\
\end{tabular}
\label{tableI3}
\end{table}

\begin{table}
\caption{Sets of tested parameters and results for \textit{fixed
    spherical grid} integration on TTF-2,5Cl$_2$BQ compound.}
\begin{tabular}{lrrrr}
\hline
\textit{Basin search} & & & &  \\
$dr_o = d_{tol}$ & 5 10$^{-3}$ & 5 10$^{-3}$ & 5 10$^{-4}$  &  5 10$^{-4}$ \\
 A & 50  & 50  & 500 & 500 \\
 B & 0.5 & 0.5 & 0.5 & 1   \\
\hline
\textit{Spherical Grid} & & & & \\
$n_ {\theta}$      & 18  & 30   & 50   & 72   \\
$n_ {\theta,\phi}$ & 406 & 1134 & 3162 & 6574 \\
$n_ {r}$           & 30  & 50   & 100  & 240  \\
\hline
CPU time & 53s & 135s & 20min. & 2.2h \\
\hline
$\delta$N$^{tot}$ & 1.534 &-0.192 &-0.204 &-0.227  \\
$\delta$N$^{val}$ & 0.147 & 0.061 & 0.041 & 0.025  \\
\hline
$\Delta$V         &-30.50 &-2.18  &-0.27  &-0.35  \\
$\sigma_{\mbox{\scriptsize CT}}$  & 12.5 & 11 & 1.2 & 1.2 \\
\hline
q$_{\mbox{\scriptsize TTF}}$         & 0.398 & 0.433 & 0.442 & 0.449  \\
q$_{\mbox{\scriptsize 2,5Cl$_2$BQ}}$ &-0.545 &-0.494 &-0.484 &-0.474 \\
\hline 
CT   & 0.47     & 0.46     & 0.46     & 0.46      \\
$\sigma_{\mbox{\scriptsize CT}}$  & 0.15 & 0.06 & 0.04 & 0.03 \\
\hline
\textbf{C$_1$} & & & &  \\
q   &-0.286  &-0.284  &-0.285  &-0.284  \\
v           & 65.14  & 65.52  & 65.59  & 65.55  \\
\hline
\textbf{S$_1$} & & & &   \\
q  & 0.251  & 0.254  & 0.256  & 0.257   \\
v           & 181.96 & 182.21 & 182.20 & 182.23  \\
\end{tabular}
\label{tableI4}
\end{table}

\begin{references}
%
\reference{Aray, Y., Rodr\'\i guez, J. \& Rivero, J. (1997)
\emph{J. Phys. Chem.} A\textbf{101}, 6976--6982.}
%
\reference{Bader, R.F.W. (1990). \textit{Atoms in Molecules: A quantum
   theory}. The international Series of Monographs on Chemistry: Oxford
   University, Clarendon Press.}
%
\reference{Bader, R.F.W. (1994). \emph{Phys. Rev. B}
   \textbf{49}-19, 13348--13356.}
%
\reference{Barzaghi, M. (2001). PAMoC (Version 2001.0), 
Online User's Manual, Centro del CNR per lo Studio delle Relazioni 
tra Struttura e Reattivit\`a Chimica, Milano, Italy, 2001. 
http://www.csrsrc.mi.cnr.it/~barz/pamoc/}
%
\reference{Biegler K\"onig, F.W., Bader, R.F.W. \& Tang, T. (1982).
\emph{J. Comp. Chem.} \textbf{3}, 317--328. AIMPAC interfaced to
GAUSSIAN, http://www.chemistry.mcmaster.ca/aimpac}
%
\reference{Biegler K\"onig, F.W.; Sch\"onbohm, J. \& Bayles, D. (2001).
\emph{J. Comp. Chem.} \textbf{22}, 545-559. AIM2000 http://www.aim2000.de/}
%
\reference{Bl\"ochl, P.E. (1994). \emph{Phys. Rev. B}
   \textbf{50}, 17953--17979.} 
%
\reference{Gatti, C., Saunders, V. R. \& Roetti, C. (1994)
\emph{J. Chem. Phys.} \textbf{101}, 10686--10696 and Gatti, C. (1996)
\emph{Acta Cryst.} A\textbf{52}, C-555--556. TOPOND interfaced to
CRYSTAL.}
%
\reference{Girlando, A., Painelli, A., Pecile, C., Calestani, G., Rizzoli, C.
\& Metzger, R.M. (1993). \emph{J. Chem. Phys.} \textbf{98}, 7692--7698.}
%
\reference{Iversen, B.B., Larsen, F.K., Souhassou, M. and Takata,
  M. (1995). \emph{Acta. Cryst.} \textbf{B51}, 580--591}
%
\reference{Katan, C. \& Koenig, C. (1999).
\emph{J. Phys.: Condens. Matter} \textbf{11}, 4163--4177.}
%
\reference{Koritsanszky, T., Howard, S., Richter, T., Su, Z., Mallinson, P. R. \&
Hansen, N. K. (1995). XD Computer Program Package for Multipolar Refinement and 
Analysis of Electron Densities from X-Ray Diffraction Data. Free University
of Berlin, Germany. Program XDPRO.}
%
\reference{Madsen, G. K. H., Gatti, C., Iversen, B. B., Damjavonic, Lj., Stucky,
G. D. \& Srdanov, V. I. (1999). \emph{Phys. Rev. B} \textbf{59}, 
12359--12369 and references therein.}
%%
\reference{Popelier, P. L. A. (1996)
\emph{Comp. Phys. Comm.} \textbf{93}, 212--240. MORPHY98
http://morphy.ch.umist.ac.uk/}
%
\reference{Popelier, P. L. A. (1998)
\emph{Comp. Phys. Comm.} \textbf{108}, 180--190}.
%
\reference{Press, W.H., Teukolsky, S.A., Vetterling, W.T. \& Flannery,
 B.P. (1992). Numerical recipes - The art of scientific
 computing. Second Edition. Cambridge University Press.}
%
\reference{Souhassou, M. \& Blessing, R. H. (1999). \emph{J. Appl. Cryst.}
    \textbf{32}, 210--217. NEWPROP interfaced to MOLLY.}
%
\reference{Stash, A. \& Tsirelson, V. (2001).
W$_{\mbox{IN}}$PRO - A Program for Calculation of the Crystal and Molecular
Properties Using the Model Electron Density, http://stash.chat.ru.}
%
\reference{Stewart, R. F. \& Spackman, M. A. (1983). VALRAY User's manual.
Carnegie-Mellon University, Pittsburgh, USA.}
%
\reference{Volkov, A. , Gatti, C.,  Abramov, Y. \& Coppens P. (2000).
\emph{Acta Cryst.} A\textbf{56}, 252--xx. Program TOPXD interfaced to XD,
http://harker.chem.buffalo.edu/public/topxd/}
\end{references}
\end{document}